\documentclass[epj,draft]{svjour}
\input{epsf}
\begin{document}
%
\title{Two interacting particles in a disordered chain II:\\
Critical statistics and maximum mixing of the one body states}
\titlerunning{TIP II: 
Critical statistics and maximum mixing of the one body states}
\author{Xavier Waintal\inst{1} \and Dietmar Weinmann\inst{2} 
\and Jean-Louis Pichard\inst{1,}\thanks{\email{pichard@spec.saclay.cea.fr}}}
\institute{CEA, Service de Physique de l'Etat Condens\'e, 
           Centre d'Etudes de Saclay, F-91191 Gif-sur-Yvette, France \and
           Institut f\"ur Physik, Universit\"at Augsburg, 
           86135 Augsburg, Germany}
\date{\today}
%
\abstract{
 For two particles in a disordered chain of length $L$ with on-site 
interaction $U$, a duality transformation maps the behavior at weak 
interaction onto the behavior at strong interaction. Around the fixed 
point of this transformation, the interaction yields a maximum mixing 
of the one body states. When $L \approx L_1$ (the one particle 
localization length), this mixing results in weak chaos accompanied 
by multifractal wave functions and critical spectral 
statistics, as in the one particle problem at the mobility edge or in certain 
pseudo-integrable billiards. In one dimension, a local interaction can only 
yield this weak chaos but can never drive the two particle system to full 
chaos with Wigner-Dyson statistics.
\PACS{{05.45.+b}{Theory and models of chaotic systems}\and 
{72.15.Rn}{Quantum localization}\and
{71.30.+h}{Metal-insulator transitions and other electronic transitions}}
}
\maketitle


 
%
%

 The competition between two body (electron-electron) interaction and 
one body kinetic energy in disordered systems is a fundamental problem 
of permanent interest. We denote by $U, t$ and $W$ the parameters 
characterizing the interaction, the one body kinetic 
energy and the fluctuations of the random potential for a 
$d$-dimensional system of size $L$. When $U$ is small, 
the $N$-body eigenstates are close to the symmetrized products of 
one body states (Slater determinants for spinless fermions)
which contain the effects of $t$ and $W$ completely. The effect 
of $U$ can be treated as a perturbation, yielding a mixing of 
those symmetrized products. When $U$ increases, the consequence of this 
mixing is that an increasing number of one body states is needed to 
describe the exact $N$-body states. If the one body states are localized 
by the disorder, delocalization in real space results from this mixing. 
This is why the interaction can induce in certain cases metallic 
behavior in a system which would be an insulator otherwise. 
When $U$ is large and dominates, one can get on the contrary a 
correlated insulator which might be metallic at weaker interaction. 
A Wigner crystal pinned by disorder is a good example of such an 
interaction-induced insulator. In the large-$U$ limit, $t$ becomes 
the small parameter, and one expands in powers of $t^2/U$. 
The issue is to know the range of validity of these 
perturbative approaches, and to describe how the system goes from 
the first limit to the second when $U$ increases. For this purpose, 
we consider a one-dimensional disordered lattice with on-site interactions.

 We discuss the simple case of two electrons with opposite spins (the orbital 
part of the wavefunction is symmetric as in the case of two bosons). 
For the main sub-band of states centered around $E=0$, both in the limits 
where $U=0$ (free bosons) and $U=\infty$ (hard-core bosons), the two body 
states can be described in terms of two one body states. We use a duality 
transformation $ U \rightarrow a t^2/U$ to map the small $U$-limit onto 
the large $U$-limit ($a \approx {\sqrt {24}}$). We first prove that the 
lifetimes of the free boson states and of the hard-core boson states are 
equal at the fixed point $U_{\rm c}$ of the duality transformation. 
At $U_{\rm c}$ one has the maximum mixing of the one body states by the 
interaction and the enhancement factor~\cite{sh} is maximum for the two 
particle localization length $L_2$. Far from $U_{\rm c}$, $L_2$ is smaller 
and satisfies the duality relation. $L_2 \rightarrow L_1$ (the one particle 
localization length) both for $U \rightarrow 0$ and $U \rightarrow \infty$. 
The study of the signature of this duality transformation on the spectral 
fluctuations is very interesting. For $E=0$, taking $L=L_1$ and increasing 
$U$, one gets two thresholds defining a range of interaction 
$U_{\rm F} \leq U \leq U_{\rm H}$. Outside this range, the levels are 
almost uncorrelated. Inside this range, the level repulsion is maximum, 
but does not reach the universal Wigner-Dyson (W-D) repulsion. The two 
particle system is not fully chaotic, but exhibits a weak chaos which 
is not arbitrarily situated between Poisson (integrable) and Wigner (chaos). 
The spacing distribution $p(s)$ between consecutive energy levels and the 
statistics $\Sigma_2(E)$ (variance of the number of energy levels inside 
an energy interval $E$) are characteristic of the third known universality 
class~\cite{sssls}. One finds $p(s)\approx 4s \exp(-2s)$ and 
$\Sigma_2(E)\approx 0.16 + 0.41 E$ for periodic boundary conditions. 
This is very close, if not identical, to the distributions found in 
many ``critical'' one body systems, such as an electron in a 3d random 
potential at the mobility edge~\cite{sssls,zk,bmp} or in certain 
pseudo-integrable quantum billiards (rational triangles~\cite{b}, 
rough billiards~\cite{fs} and  Kepler problem~\cite{al}). Furthermore, 
$p(s)$ saturates to $4s \exp(-2s)$ for $U \approx U_{\rm c}$ 
only when the ratio $1 \leq L_1/L \leq 10$. We 
conclude that a local interaction can never drive the two particle 
system to full quantum chaos with Wigner-Dyson statistics in one dimension, 
but can at most yield weak critical chaos in a certain domain of interaction 
and of the ratios $L_1/L$. We show in addition that this weak chaos is 
accompanied by multifractal wavefunctions, in agreement with Ref.~\cite{wp}.

 Each one particle Hamiltonian is given by 
\begin{equation}
H_0 = t \sum_{n=1}^{L-1} (|n\rangle\langle n+1| + |n+1\rangle\langle n|)+ 
 \sum_ {n=1}^{L}V_n |n\rangle\langle n|
\end{equation}
where $V_n$ is uniformly distributed inside $[-W,+W]$ and the interaction 
is described by 
\begin{equation}
H_{\rm int} = U  \sum_{n=1}^L |n n\rangle\langle n n|.
\end{equation}
%

  Denoting $|\alpha\rangle$ the one body state of energy $\epsilon_{\alpha}$  
and the amplitude $\langle n|\alpha\rangle= \Psi_{\alpha}(n) $, only two 
one body states $|\alpha\rangle $ and $|\beta\rangle $
are necessary to describe a free boson state $|fb\rangle =|\alpha\beta\rangle$
with components $\langle n_1n_2|fb\rangle$ given by 
\begin{equation}
 ( \Psi_{\alpha}(n_2) \Psi_{\beta}(n_1) + 
\Psi_{\alpha}(n_1) \Psi_{\beta}(n_2))/\sqrt{2}. 
\end{equation}
In this free boson basis, the interaction matrix elements 
 $\langle\alpha\beta | H_{\rm int} | \gamma\delta \rangle 
 = 2U Q_{\alpha\beta}^{\gamma\delta} $ 
where   
\begin{equation}
Q_{\alpha\beta}^{\gamma\delta}= \sum_{n=1}^L  \Psi_{\alpha}(n) 
\Psi_{\beta}(n) \Psi_{\gamma}(n) \Psi_{\delta}(n) .
\end{equation} 
 This $Q$-matrix has been studied~\cite{wp} in one dimension for $L \geq 
L_1$. It was found that its lines, where are the terms 
$Q_{\alpha\beta}^{\gamma\delta}$ coupling a given $|\alpha\beta\rangle$ 
to all the other $|\gamma\delta\rangle$, is a multifractal measure in the 
$|\gamma\delta\rangle$ space. Therefore the effective density 
$\rho_2^{\rm eff}$ of $|fb\rangle$ states coupled by the square of the 
interaction matrix elements is reduced. For $L \approx L_1$, one has 
$\rho_2^{\rm eff} (L_1) \propto L_1^{1.75}$, instead of the total two 
body density $\rho_2 \propto L_1^2$. This was confirmed by a study of 
their Fermi golden rule decay.  


 As noticed in Ref.~\cite{si}, there is a useful representation 
for an on-site interaction $U \rightarrow \pm \infty$, composed
 by a small set of $L_{\rm M}=L$ ``molecular'' states $|nn\rangle$ and 
by a large set of $L_{\rm H}=L(L-1)/2$ hard core boson states $|hc\rangle$ 
 built from re-symmetrized Slater determinants. Their components 
$\langle n_1 n_2|hc\rangle $ are given by 
\begin{equation}
\frac {(\Psi_{\alpha}(n_2) 
\Psi_{\beta}(n_1) - \Psi_{\alpha}(n_1) \Psi_{\beta}(n_2)) 
{\rm sgn}(n_2 - n_1)}{\sqrt{2}}. 
\end{equation}
The re-symmetrization is insured by the function ${\rm sgn}(x):= x/|x|$.  
For hard wall boundaries (which we assume in our discussion), one has to 
use the same one body states $|\alpha\rangle$ and $|\beta\rangle$ both for 
$|fb\rangle$ and $|hc\rangle$. However, if the system is closed on itself and 
forms a ring threaded by a flux $\Phi$, the re-symmetrization on a torus 
reveals interesting topological aspects (see ref.~\cite{ap}). 
One has to use different one body states in the two limits which differ
in $\Phi$ by half a flux quantum (to periodic boundary conditions for $U=0$ 
correspond anti-periodic boundary conditions for $U\rightarrow \pm \infty$). 

In this basis, the two body Hamiltonian ${\cal H}$ has the structure
 \begin{equation} 
 {\cal H}= \left[
 \begin{array}{cc}
  H_{\rm M}  & H_{\rm C}  \\
  H_{\rm C}^T & H_{\rm H} \\
 \end{array}
 \right] \, .
\end{equation}
$H_{\rm M}$ and $H_{\rm H}$ are 
 $L_{\rm M} \times L_{\rm M}$ 
and $L_{\rm H}\times L_{\rm H}$
diagonal matrices with entries $U+2V_n$ and $\epsilon_{hc}=\epsilon_{\alpha}+
 \epsilon_{\beta}$, respectively. $H_{\rm M}$ and $H_{\rm H}$ are coupled 
by a rectangular matrix $H_{\rm C}$ resulting from the kinetic terms 
 $\sqrt{2}\, t \sum_{n=1}^{L-1}(|n,n+1\rangle\langle n,n| + 
|n,n\rangle\langle n,n+1| +{\rm h.c.}) $ of ${\cal H}$. The matrix elements 
of $H_{\rm C}$ are equal to 
 $\sqrt{2}\, t \left (\Psi_{\beta}(n) \hat{D} \Psi_{\alpha}(n)
 -\Psi_{\alpha}(n) \hat{D} \Psi_{\beta}(n) \right )$, where 
 $\hat{D} \Psi_{\alpha} (n) := \Psi_{\alpha}(n+1) - \Psi_{\alpha}(n-1)$. 
The states $|hc\rangle$ of energy $\epsilon_{hc}\approx 0$ are coupled by 
a term of order $t$ to the few states $|nn\rangle$ of energy $\approx U$. 
Their lifetime becomes infinite when $U \rightarrow \pm \infty$. 

\begin{figure}[tb]
\epsfxsize=3in
\epsfysize=3in
\epsffile{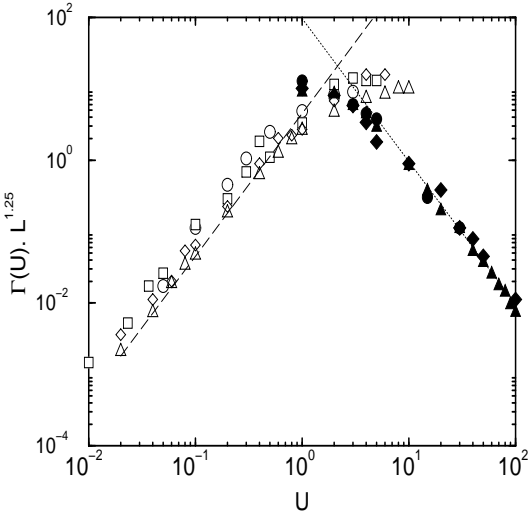}
\vspace{3mm}
\caption[fig1]{\label{fig1} $\Gamma_0$ (open symbols) and $\Gamma_{\infty}$ 
(filled symbols) for $L=25$ (triangles), $L=76$ (diamonds), $L=150$ (squares) 
and $L=200$ (circles). Dashed (dotted) lines are the Fermi golden rule 
expression of Eq.\ref{fgr1} (Eq.\ref{fgr2}).}
\end{figure}

 Projecting an eigenstate $|A\rangle$ of energy 
 $E_A$ onto the states $|nn\rangle$ and $|hc\rangle$ 
 \begin{equation}
 |A\rangle =\sum_{n=1}^{L_{\rm M}} c_A^n |nn\rangle + 
 \sum_{hc=1}^{L_{\rm H}} c_A^{hc} 
 |hc\rangle \, ,
 \end{equation}
 one finds the relation 
 $$\left[
 \begin{array}{cc}
 H_{\rm M} + J_{\rm M} (E_A) & 0 \\
 0 & H_{\rm H} + J_{\rm H} (E_A) \\
 \end{array}
 \right] 
 \left[
 \begin{array}{cc}
 C_A^{\rm M} \\
 C_A^{\rm H} \\
 \end{array}
 \right] 
 = E_A
 \left[
 \begin{array}{cc}
 C_A^{\rm M} \\
 C_A^{\rm H} \\
 \end{array}
 \right],$$
 which holds for arbitrary $U$. $C_A^{\rm M}$ and $C_A^{\rm H}$ are vectors 
 of $L_{\rm M}$ components $c_A^n$ and $L_{\rm H}$ components $c_A^{hc}$ 
 respectively. The matrix $J_{\rm M}(E_A)$ has $L_{\rm M}^2$ elements 
 given by 
\begin{equation}
J_{\rm M}(E_A,nn,mm) = \sum_{hc=1}^{L_{\rm H}} \frac 
{(\langle nn|H_{\rm C}|hc\rangle\langle hc|H_{\rm C}|mm\rangle)}
{E_A-\epsilon_{hc}}
\end{equation}
 and the matrix $J_{\rm H}(E_A)$ has $L_{\rm H}^2$ elements of the 
 form 
\begin{equation}
J_{\rm H}(E_A,hc,\tilde{hc})=\sum_{n=1}^{L_{\rm M}} \frac 
{\langle hc|H_{\rm C}|nn\rangle\langle nn|H_{\rm C}|{\tilde{hc}}\rangle}
 {U+2V_n-E_A}.
\end{equation}
 We consider the main sub-band of states with energy $E_A\approx 0$, 
resulting from the mixing by a perturbation of order $t^2/U$ of a few 
states $|hc\rangle$ for which $J_{\rm H}(E_A)$ can be simplified. 
 Assuming $W,t \ll U$ and $U+2V_n-E_A \approx U$, one has to evaluate 
$\sum_{n=1}^L 
\langle hc|H_{\rm C}|nn\rangle\langle nn|H_{\rm C}|{\tilde{hc}}\rangle$. 
This expression is composed of 12 sums over $n$, each of them having the 
form 
 \begin{equation}
 \tilde{Q}_{\alpha\beta}^{\gamma\delta}=\sum_{n=1}^L  \Psi_{\alpha}(n) 
 \Psi_{\beta}(n') \Psi_{\gamma}(n) \Psi_{\delta}(n'') 
 \end{equation} 
with various combinations of $n',n'' = n\pm 1$. Therefore 
$\tilde{Q}_{\alpha\beta}^{\gamma\delta}$ is not exactly equal to the 
$Q_{\alpha\beta}^{\gamma\delta}$ occuring around the free boson 
limit. However, this difference should not be statistically relevant 
and if one neglects it, one finds 
\begin{equation}
J_{\rm H}(E_A,hc,\tilde{hc}) \approx \pm 2(24)^{1/2} {t^2 \over U} 
Q_{\alpha\beta}^{\gamma\delta}.
\end{equation}
%
%
 This establishes the duality transformation $U \rightarrow \sqrt{24} 
(t^2/U)$ for $E \approx 0$ which maps the distribution of the coupling 
terms between the $|fb\rangle$ when $U$ is small onto the distribution 
of the coupling terms between the $|hc\rangle$ when $U$ is large. 
To illustrate this duality around the fixed point $U_{\rm c} = (24)^{1/4} 
t$, we have numerically calculated the average over the disorder of the 
local density of states
\begin{equation}
\rho_{A} (E) = \sum_{\alpha\beta} |c_A^{\alpha\beta}|^2 
 \delta (E+E_A-\epsilon_{\alpha}-\epsilon_{\beta})
\end{equation}
for $L=L_1$ and $E_A \approx 0$. We have observed that 
$\langle\rho_A(E)\rangle$ can be fitted by a Lorentzian curve of width 
$\Gamma_0$ (if  $\alpha\beta$ denotes the states $|fb\rangle$) 
or $\Gamma_{\infty}$ (if $\alpha\beta$ denotes the states $|hc\rangle$).
 Fermi's golden rule gives for the widths $\Gamma$ at $L\approx L_1$
\begin{eqnarray}
\label{fgr1}
\Gamma_0 &\approx& 2\pi U^2 {1\over L_1^3} \rho_2^{\rm eff} (L_1) \\
\label{fgr2}
\Gamma_{\infty} &\approx & 2 \pi \frac{24 t^4}{ U^2} 
\frac{1}{L_1^3} \rho_2^{\rm eff} (L_1) 
\end{eqnarray}
where $\rho_2^{\rm eff} \approx L_1^{1.75} / t \neq  \rho_2 (L_1) 
\approx L_1^2/t$ because of the multifractality of the $Q$-matrix. 
As shown in figure~\ref{fig1}, the widths $\Gamma$ obey the duality 
property $\Gamma_0 (U) = \Gamma_{\infty} (\sqrt{24} t^2/U)$, and are 
described by the above golden rule approximations. When 
$U \approx U_{\rm c}$, the lifetime of the free boson states is equal 
to the one of  the hard core boson states. 

%
%

   When $U$ increases, the statistical ensemble associated to the 
two particle system exhibits a crossover from one preferential basis 
(the free boson basis) to another preferential basis (the hard core 
boson basis). At $U_{\rm c}$, the statistical ensemble is in the middle 
between the two preferential bases, the mixing of the one body states is 
maximum, and the localization length $L_2$ therefore is maximum. 
When $E=0$ and $U$ varies, a transition occurs in the two body 
 spectrum in 1d which is somewhat reminiscent of the one body case in 3d 
when $E=0$ and $W$ varies. In the two cases, one can expand in powers 
of $U/t$ or $t/U$, of $W/t$ or $t/W$ when the system 
 is under or above the critical threshold, respectively. The analogy is 
of course not complete, since the two body spectrum in 1d has essentially 
Poisson statistics in the two perturbative limits, while the one body 
spectrum in 3d exhibits W-D statistics for $W \ll t$ and Poisson statistics 
when $W \gg t$. Nevertheless, the question whether or not the spectral 
fluctuations are of the same kind in the vicinity of the threshold deserves 
to be investigated. 

 Before showing the results, two arguments can be mentioned: (i) A 
multifractal $Q$-matrix is incompatible with Wigner-Dyson level repulsion 
in 1d. The states $|fb\rangle$ or $|hc\rangle$ are directly coupled by $U$ or 
$t^2/U$ to an effective density $\rho_2^{\rm eff}<\rho_2$. Therefore, 
nearest neighbors in energy are very likely uncorrelated, enhancing the 
probability to find level spacings small compared to their average. (ii) 
Looking at Eq.\ (2), one may assume that a broad distribution of the 
matrix elements of $H_{\rm M}$ may yield a single dominant coupling term. 
This allows us to consider that the states $|hc\rangle$ are mainly coupled 
via a few states $|nn\rangle$. This is not far from the case discussed 
by Bohr and Mottelson~\cite{bm} (coupling via a single state) where the 
consecutive levels $E_A$ of ${\cal H}$ satisfy the conditions 
$\epsilon_{hc} < E_A < \epsilon_{hc+1} < E_{A+1}$, $\epsilon_{hc}$ being 
the consecutive levels of $H_{\rm H}$. Since the statistics of $H_{\rm H}$ 
is essentially Poissonian, this forbids to have the Wigner-Dyson rigidity 
for the $E_A$. The most rigid spectrum would be achieved by putting the $E_A$ 
exactly in the middle of two consecutive $\epsilon_{hc}$. It is 
straightforward to find that $p(s)$ would then be equal to 
$p_{\rm c}(s)=4s \exp(-2s)$. This ``semi-Poisson'' distribution, where 
$p_{\rm c}(s) \propto s$ for $s\ll 1$ (as the Wigner surmise 
$p_{\rm W}(s)= (\pi/2) s \exp(-\pi s^2/4)$), and which decays as $\exp(-s)$ 
for $s\gg 1$ (as the Poisson distribution  $p_{\rm P}(s)$), is attracting
increasing interest since it characterizes~\cite{b} weak chaos in systems 
which are ``critical'' (i.\ e.\ which can be mapped~\cite{al} in a certain 
``integrable'' basis onto an Anderson model at its critical point). It was 
shown in Ref.~\cite{jpb} that a plasma model for the energy levels can 
reproduce the level statistics of a disordered conductor if the logarithmic 
Wigner-Dyson level repulsion was screened at an energy scale given by the 
Thouless energy $E_T$. Since $E_T$ is of order of the mean level spacing 
$\Delta_1$ at the mobility edge, one can expect that a short range plasma 
model should also give this ``semi-Poisson'' distribution. This was recently 
confirmed~\cite{b} for a logarithmic level repulsion truncated to the 
first consecutive level, where one finds $p(s)=4s \exp(-2s)$ and 
$\Sigma_2 (E) = 1/16 + E/2$. After averaging over certain boundary 
conditions, $p_{\rm c}(s)$ was also found to 
give~\cite{bmp} a good fit for the $p(s)$ of the one body spectrum in 
3d at the mobility edge.
\begin{figure}[tb]
\epsfxsize=3in
\epsfysize=3in
\epsffile{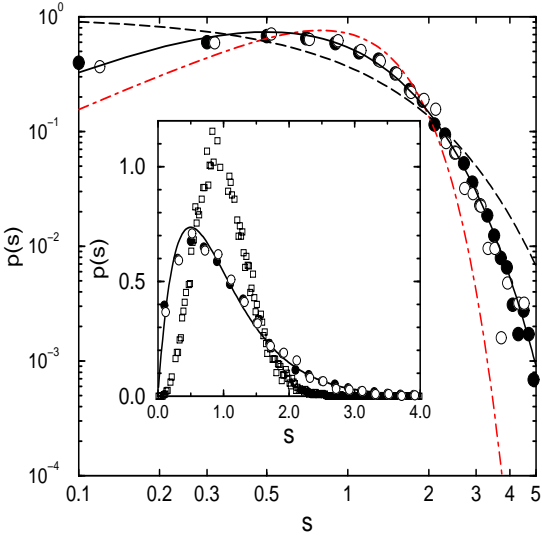}
\vspace{3mm}
\caption[fig2]{\label{fig2} 
 TIP-$p(s)$ for $L=L_1=150$ and $U=1$: Continuous, dotted and dashed lines 
correspond to the ``Semi-Poisson'', Wigner and Poisson distributions 
respectively. Full (empty) circles correspond to hardwall (periodic) 
boundary conditions. Inset: TIP-$p(s)$ (circles $U=1$) compared to the 
one particle $p(s)$ (squares) for $L=L_1=150$.}
\end{figure}

\begin{figure}[tb]
\epsfxsize=3in
\epsfysize=3in
\epsffile{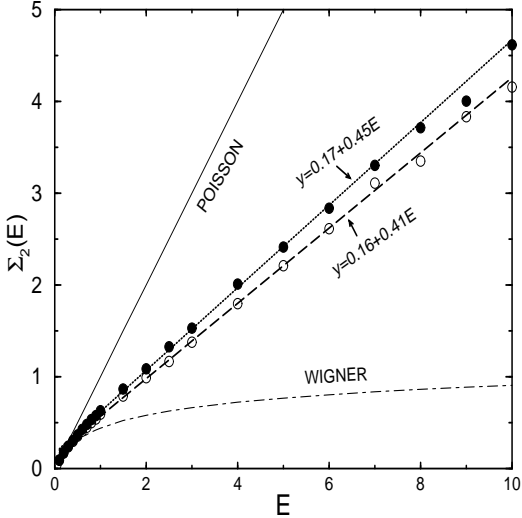}
\vspace{3mm}
\caption[fig3]{\label{fig3}
$\Sigma_2(E)$  for $L=L_1=100$ and $U=1.25$ $E$ in units of the 
mean level spacing. Full (empty) circles correspond to hardwall (periodic) 
boundary conditions.}
\end{figure}

 These observations lead us to study $p(s)$ as a 
function of $U$ and of the ratio $L_1/L$ around $E=0$, where $\rho_2$ has 
a van Hove singularity for $W=U=0$. The disorder and the interaction 
remove the singularity and $\rho_2$ develops a small plateau around 
$E\approx 0$. It is inside this plateau that $p(s)$ and $\Sigma_2$ have been 
investigated.  For $U=1$ and $L=L_1=150$, the spacing distribution $p(s)$ is 
shown in Fig.~\ref{fig2}, in good agreement with the ``semi-Poisson'' 
distribution. In the inset, one can see that this is specific to the 
two particle problem, and does not characterize the single particle spectrum 
for the same chain of size $L_1$. 
 
  The statistics $\Sigma_2(E)$ displays also the behavior that one 
expects at a critical point or for a short range plasma model. 
$\Sigma_2 (E) \approx \chi_0 + \chi_1 E$ is shown in 
fig.~\ref{fig3}. The values of the compressibility 
$\chi_1=\lim_{E\rightarrow \infty} \Sigma_2(E)/E$  seem to depend on 
the boundary conditions for the first consecutive levels, as 
for the Anderson transition in 3d. The values for the two particle 
system $0.41$ (periodic) and $0.45$ (hard wall) coincide with the 
values given~\cite{bmp} by Braun et al. for certain combinations of 
boundary conditions in the different directions. This 
sensitivity of $\chi_1$ to the boundary conditions may disappear 
for larger $E$ (smaller time scales), as it is the case for the one 
particle spectrum at the mobility edge~\cite{mont}. 

 We now study the domain of validity for weak chaos and universal 
critical statistics.  To measure the deviation of $p(s)$ from 
the $P_{\rm P}(s)$ or $P_{\rm W}(s)$, we use the functional 
\begin{equation}
\eta(U)=\frac{\int_0^b{\rm d}s(p(s)-
p_{\rm W}(s))}{\int_0^b{\rm d}s(p_{\rm P}(s)-p_{\rm W}(s))}
\end{equation}
with $b=0.4729$. A Poisson spectrum gives $\eta = 1$, a 
Wigner-Dyson spectrum gives $\eta=0$ and ``semi-Poisson'' corresponds 
to $\eta_{c}=0.386$. 

\begin{figure}[tb]
\epsfxsize=3in
\epsfysize=3in
\epsffile{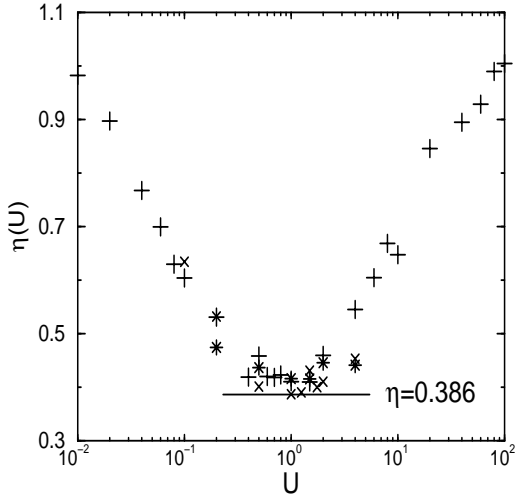}
\vspace{3mm}
\caption[fig4]{\label{fig4} Weak chaos and interaction for $L=L_1$: 
Crosses, pluses and stars represent $\eta(U)$ for $L=100$, $L=150$ and 
$L=200$, respectively.}
\end{figure}

First, we vary $U$, imposing the relation 
$L=L_1$. We assume $L_1$ to be given by the weak disorder formula 
$L_1\approx 25/W^2$. In the limits $U \rightarrow 0$  and $ U \rightarrow 
\infty$, the TIP-levels are given by $\epsilon_{bf}=\epsilon_{hc}=
\epsilon_{\alpha}+\epsilon_{\beta}$ which are uncorrelated (at least on 
energy scales smaller than the one particle level spacing $\Delta_1$). 
This gives $p_{\rm P}(s)$ for the TIP-spectrum. 
One can see in fig.~\ref{fig4} that $p(s)$ deviates for small $U$ and 
$t^2/U$ from $p_{\rm P}(s)$ observed at $U=0$ and $U=\infty$. $\eta(U)$ 
exhibits a plateau ($U_{\rm F} < U < U_{\rm H}$) at the value 
$\eta_{\rm c}\approx 0.386$ which characterizes $p_{\rm c}(s)$). We 
suggest that $U_{\rm F}$ and $U_{\rm H}$ are given by the conditions 
which hold~\cite{wpi} for systems in which there is no coupling 
term between consecutive energy levels. In agreement with the general 
picture developed in Ref.~\cite{wpi} (see also Ref.~\cite{js2}), the 
threshold appears when the strength of the coupling terms becomes of 
the order of the spacing $1/\rho_2^{\rm eff}$ of the directly coupled 
levels. This gives $2U_{\rm F}/L_1^{3/2} \approx 4t/ L_1^{1.75}$ and 
$2 \sqrt{24} t^2/(U_{\rm H} L_1^{3/2}) \approx 4t/L_1^{1.75}$, 
respectively, and seems to account for the size of the plateau of the 
curve $\eta(U)$.  This favours a line of critical points rather than 
an isolated point when the condition $L=L_1$ is imposed. The 
$U$-dependence of $\chi_1$ exhibits the same plateau as $\eta(U)$. 
The results shown in fig.~\ref{fig3} for $U=1.25$ do not vary inside 
the plateau.

\begin{figure}[tb]
\epsfxsize=3in
\epsfysize=3in
\epsffile{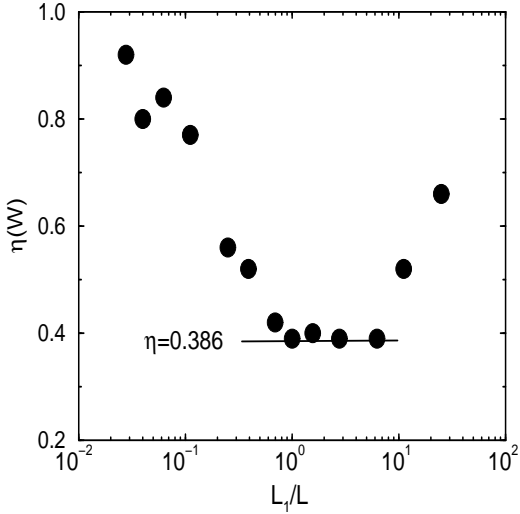}
\vspace{3mm}
\caption[fig5]{\label{fig5} 
Weak chaos and disorder for $U=1.25$: $\eta(W)$ as a function of $L_1/L$ 
with $L=100$.}
\end{figure}

 Second, for $U=1.25$ and $L=100$, we vary the disorder parameter 
$W$ in order to study the role of the ratio $L_1/L$. When $L \gg L_1$, 
the small fraction of TIP-states re-organized by $U$ 
disappears behind the set of TIP-states which are not re-organized 
by $U$ and which remain uncorrelated. $p(s)$ does not differ from 
$p_{\rm P}(s)$. On the contrary, when $L \ll L_1$, we approach 
the clean limit which can be easily understood: For $W=0$, the total 
TIP-momentum $K$ is conserved and the Hamiltonian matrix 
has a block diagonal form in the free boson eigenbasis. Denoting by 
$\epsilon_{\gamma}  = 2\cos k_{\gamma}$ the eigenenergy  of a single 
particle with momentum $k_{\gamma}$, and assuming periodic boundary 
conditions, one can easily show that the $(L+1)/2$ TIP eigenenergies 
$E_A (K)$ of total momentum $K$ modulo ($2\pi$) satisfy 
 \begin{equation}
 \label{eq(W=0)}
 \frac{2U}{L} \sum_{k_{\gamma}, k_{\delta}} \frac {1}{E_A (K) 
 -\epsilon_{\gamma}  -\epsilon_{\delta}}=1.
 \end{equation}
 The summation is restricted to the momenta $k_{\gamma} + k_{\delta} 
 = K$ modulo $(2\pi$). This sequence of $(L+1)/2$ levels 
 exhibits the same property than in the case discussed by 
 Bohr and Mottelson: For a given $K$, the levels for $U\neq 0$ 
 alternate with the levels for $U=0$.  This sequence of $E_A (K)$ can only 
 exhibit level repulsion between next 
 nearest neighbors, but not the long-range repulsion necessary for 
 Wigner-Dyson statistics. Since the total spectrum is the uncorrelated 
 superposition of the $L$ different sequences associated to different total 
 momenta $K$, one gets for the TIP spectrum uncorrelated levels for any 
 value of $U$ when $W=0$. 

 Between the two limits $L \ll L_1$ and $L \gg L_1$, one can see in 
 fig.~\ref{fig4} a plateau where the spectrum is more rigid, $\eta (L_1/L)$ 
 saturating to the ``semi-Poisson'' value. This plateau for $U=1.25$ is 
 observed for $1 \leq L_1/L \leq 10$. 

\begin{figure}[tb]
\epsfxsize=3in
\epsfysize=2.7in
\epsffile{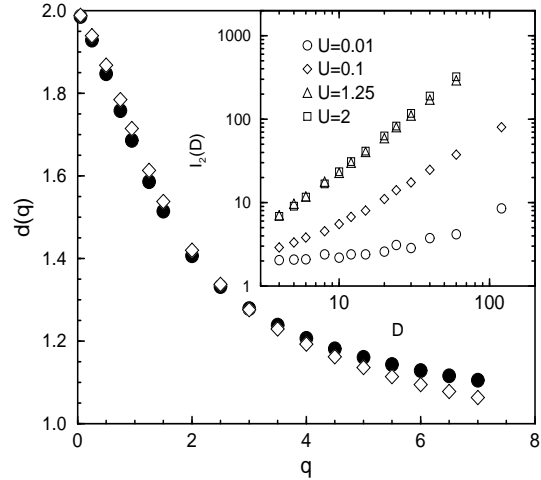}
\vspace{3mm}
\caption[fig6]{\label{fig6} $d(q)$ at $U=2$ for $L=L_1=120$ in the 
bases $|fb\rangle$ (open symbols) and $|hc\rangle$ (filled symbols). 
The inset shows the establishment of the power-law behaviour of 
$I_2(D)$ for increasing values of $U$ in the $|hc\rangle$ basis.}
\end{figure}

%
%

 A critical statistics for the spectrum is usually related to multifractal 
wave functions. We have studied the projection $C_A^{\alpha\beta}$ of an 
eigenstate $|A\rangle$ onto the states $|fb\rangle$ and onto the states 
$|hc\rangle$. 
Since the perturbation matrix elements (divided by $U$ 
or $t^2/U$) are multifractal, one expects that this will eventually 
yield a multifractal structure for $|A\rangle$ in the two bases for 
large and small enough $U$, respectively. 
We have done the same analysis  as in Ref.~\cite{wp} to which we refer for 
technical details and references. We divide the plane $(\alpha,\beta)$ 
into $(L/D)^2$ boxes of size $D$ and compute for each box $i$ the probability 
$p_i=\sum_{\alpha\beta \in {\rm box_i}}|C_A^{\alpha\beta}|$. After ensemble 
averaging, the moments $I_q(D)=\sum_i p_i^q$ have a power-law behaviour if 
$U$ is large enough for the $|fb\rangle$ basis (inset of fig.~\ref{fig3}), 
and if $U$ is small enough for the $|hc\rangle$ basis. So one finds 
$I_q(D) \propto D^{(q-1) d(q)}$, $d(q)$ defining the corresponding 
generalized Renyi dimensions. The $d(q)$ are roughly equal in the two 
bases (fig.~\ref{fig3}) and do not depend on $U$ as soon as the 
power-law behaviour is established for $I_q(D)$, i.\ e.\ inside the 
line of fixed points.

In summary, we have shown that this simple two particle system in 
1d is characterized by two perturbative regimes (in $U$ or $t^2/U$). 
These regimes are dominated by the existence of a single preferential 
basis and are related to each other by a duality transformation. 
In the middle, the mixing of the one body states by the interaction 
is maximum, but the system is not chaotic (no Wigner-Dyson statistics). It 
exhibits the weak chaos of the critical systems, as confirmed by the 
spectral fluctuations. The curves $\eta(U,W), \chi_1(U,W)$ and $d_q(U,W)$ 
in the $|fb\rangle$ and $|hc\rangle$ bases are consistent with the existence 
of a weak critical chaos for a given size in $(U,W)$ domain located around 
$U_{\rm c}$ and $L_1/L\approx 1$. In one dimension, a local interaction can 
never drive a two particle system towards full chaos with Wigner-Dyson 
statistics. 

\begin{acknowledgement}
 We acknowledge useful discussions with E.\ Bogomolny, G.\ Montambaux 
and P.\ Silvestrov. 
\end{acknowledgement}
%
%



\end{document}